\documentclass[3p,times]{elsarticle}

\usepackage{xspace}
\usepackage[T1]{fontenc}
\usepackage[scaled]{beramono}
\usepackage{listings}
\usepackage{tikz}
\usepackage{dingbat}
\usepackage{amssymb}
\usepackage{amsmath}
\usepackage{bbding}
\usepackage{pbox}
\usepackage{scrextend}
\usepackage{hyperref}

\usetikzlibrary{shapes,arrows,chains}

\pgfdeclarelayer{marx}
\pgfsetlayers{main,marx}
\providecommand{\cmark}[2][]{\relax}

\newcommand{\hope}{\texttt{HOPE}\xspace}

\begin{document}

\begin{frontmatter}

\title{HOPE: A Python Just-In-Time compiler for astrophysical computations}

\author{Jo\"el Akeret\corref{cor1}}
\ead{jakeret@phys.ethz.ch}

\author{Lukas Gamper}
\author{Adam Amara}
\author{Alexandre Refregier}

\cortext[cor1]{Corresponding author}

\address{ETH Zurich, Institute for Astronomy, Department of Physics, Wolfgang Pauli Strasse 27, 8093 Zurich, Switzerland}

\begin{abstract}
The {\tt Python} programming language is becoming increasingly popular for scientific applications due to its simplicity, versatility, and the broad range of its libraries. A drawback of this dynamic language, however, is its low runtime performance which limits its applicability for large simulations and for the analysis of large data sets, as is common in astrophysics and cosmology. While various frameworks have been developed to address this limitation, most focus on covering the complete language set, and either force the user to alter the code or are not able to reach the full speed of an optimised native compiled language.  In order to combine the ease of {\tt Python} and the speed of {\tt C++}, we developed \hope, a specialised {\tt Python} just-in-time (JIT) compiler designed for numerical astrophysical applications. \hope focuses on a subset of the language and is able to translate {\tt Python} code into {\tt C++} while performing numerical optimisation on mathematical expressions at runtime. To enable the JIT compilation, the user only needs to add a decorator to the function definition. We assess the performance of \hope by performing a series of benchmarks and compare its execution speed with that of plain {\tt Python}, {\tt C++} and the other existing frameworks. We find that \hope improves the performance compared to plain {\tt Python} by a factor of 2 to 120, achieves speeds comparable to that of {\tt C++}, and often exceeds the speed of the existing solutions. We discuss the differences between \hope and the other frameworks, as well as future extensions of its capabilities. The fully documented \hope package is available at \url{http://hope.phys.ethz.ch} and is published under the GPLv3 license on PyPI and GitHub.
\end{abstract}


\end{frontmatter}

\section{Introduction}
\label{sec:introduction}

In recent years, the {\tt Python} programming language has gained a wide acceptance beyond its original use of simple scripting for system administration and test automatisation. {\tt Python} has evolved to become a primary programming language for various industries and many scientific research fields, notably in astrophysics \cite{tiobe2014,ieee2014}. The reasons for this success are the simplicity, flexibility and maintainability of {\tt Python} code as well the wide range of its libraries. In particular, the widely used numerical {\tt Python} packages {\tt NumPy}\cite{numpy} and {\tt SciPy}\cite{scipy} allow for fast prototyping and development of new applications. The flexible runtime and the dynamic typing are further features of the interpreter language that are valued by developers. However, these advantages come with a drawback: typically the execution time of {\tt Python} programs can be slower than native compiled languages such as {\tt C} or {\tt Fortran} by orders of magnitudes.

While for many applications the performance of the software is not a priority, in astrophysics and cosmology where large simulations and data analysis over large data sets are often required, speed can be crucial (see e.g. \cite{refregier2013, akeret2013} and references therein for cosmology). In order, to increase the overall performance one can parallelize the programs to take advantage of multicore CPU architectures. An alternative and complementary approach is to focus on improving the single thread performance. The problem of improving the single thread performance of dynamic languages, such as Python, has been addressed in various difference ways that can be grouped broadly into two categories: 1) development of optimizing just-in-time (JIT) compilers and 2) development of faster interpreters  \cite{Arnold2005}. The concept of the latter is to reduce the overhead introduced by the dynamic nature of the language. The idea of a JIT compiler is to produce faster machine or byte code during runtime when needed \cite{antocuni_phd_2010}. In the {\tt Python} landscape both approaches have been implemented in various projects e.g. the {\tt PyPy}\footnote{\url{http://www.pypy.org}} interpreter, the {\tt Numba}\footnote{\url{http://numba.pydata.org}} JIT package or the {\tt Cython}\footnote{\url{http://www.cython.org}} C-extension and others.

While most of these approaches aim to support the complete language or a large portion of it, we find that some of the solutions are often intrusive (i.e. require the user to tailor the code to the framework) and that only a few of them are able to reach the full speed of an optimized native compiled {\tt C++} code. To fully combine the ease of {\tt Python} and the speed of C++, we have therefore developed the \hope package. \hope is a specialized {\tt Python} JIT compiler that supports a subset of the {\tt Python} language - primarily numerical features commonly used in astrophysical calculations - and aims to reach the highest possible execution speed. The package translates {\tt Python} code into {\tt C++} and is able to perform numerical optimization on mathematical expression at runtime. By using \hope, the user benefits from being able to write common numerical code in {\tt Python} while having the performance of compiled implementations. To enable the \hope JIT compilation, the user only needs to add a decorator to the function definition. The package does not require additional information, which ensures that \hope is as non-intrusive as possible. 

We used the \hope package in astrophysics applications such as the development of {\tt PyCosmo} \cite{refregier2014} and to rewrite {\tt UFig} (the Ultra fast image generator) \cite{berge2013}  into {\tt Python}. {\tt PyCosmo} is a {\tt Python} cosmology package that numerically integrates the Einstein-Boltzmann differential equations and computes various cosmological observables. The \hope package allowed us to improve the performance of the integration by a factor of ~50 $\times$ compared to the pure {\tt Python} implementation by JIT compiling the integrated function. {\tt UFig} is an image generator developed to simulate wide field cosmological galaxy surveys thus enabling tight control of systematics effects through forward modelling \cite{refregier2013, berge2013}. \hope has allowed us to rewrite the original {\tt C++} implementation of {\tt UFig} into {\tt Python}. Benchmarks show a performance comparable to the earlier {\tt C++} code, but the new implementation has increased the modularity, extensibility and readability of the code. 

This paper is organised as follows. In section \ref{sec:frameworks}, we discuss the reason for the lower execution speed of dynamic languages, and review the existing solutions to address the performance implications in {\tt Python}. Section \ref{sec:architecture} explains how we address the performance requirements using our JIT compiler package and describe the \hope architecture and design. In section \ref{sec:benchmarks} we introduce the benchmarks we used to compare the performance of \hope to the existing solutions. We show and discuss results in sections \ref{sec:results} and \ref{sec:discussion} and finally conclude in section \ref{sec:conclusion}. Information for downloading the \hope package and on its performance on an alternative platform are described in Appendix A and B, respectively.

\section{Review of existing solutions}
\label{sec:frameworks}

{\tt Python} is a dynamic interpreted language, which requires an interpreter for the execution of a program. Typically the performance of {\tt Python} is much slower than {\tt C} or comparable compiled languages. There are several reasons for this: {\tt Python} has not been designed to be a fast language, but, instead, readability has been defined to be more important. In addition, everything in {\tt Python}, including simple values, are represented as objects in memory. Therefore, the evaluation of a numerical expression requires the interpreter to unbox every value and to load the appropriate operation, as operators can be overwritten dynamically.

Currently four main implementation of interpreters exist: {\tt CPython}, {\tt Jython}, {\tt IronPython} and {\tt PyPy}. {\tt CPython} is the reference implementation written in {\tt C} and the most widely used {\tt Python} interpreter. {\tt Jython} and {\tt IronPython} are alternative implementations of {\tt Python} interpreters targeting the Java virtual machine and the .NET frameworks, but they typically do not improve performance\cite{antocuni_phd_2010}. {\tt PyPy} is the most prominent effort to develop an alternative interpreter in order to increase the execution speed. In the following, we always refer to the {\tt CPython} implementation if not explicitly stated otherwise.

If performance is critical, one approach is to write the critical code in {\tt C} or {\tt  C++} and interface the implementation with {\tt Python} bindings. With this approach the performance requirements can typically be met but often at the cost of the readability. Experience shows that this makes it difficult for diverse users to maintain and extend the {\tt C/C++} code. A further drawback of this approach is that every change in the code requires a recompilation. During development this has to be performed repeatedly, which lowers the productivity and the development velocity of the team.

To be able to implement a code in {\tt Python} but nevertheless be able to properly address the performance needs, different alternative approaches exist, including: {\tt Numba}, {\tt Cython}, {\tt Nuitka} and {\tt numexpr} (see Table \ref{tbl:platforms}). The {\tt Numba} package is a JIT compiler that generates optimized machine code at runtime using the {\tt LLVM} compiler infrastructure. A similar but different approach is to translate and compile the {\tt Python} source code. The {\tt Cython} C-extension for {\tt Python} is widely used for this. The {\tt Nuitka} project\footnote{\url{http://www.nuitka.net}} is a static {\tt Python} compiler that focuses on full support of the language. A further approach is the {\tt numexpr} package\footnote{\url{https://github.com/pydata/numexpr}} that specialises in evaluating numerical expressions and additionally allows parallel execution on multiple cores. Beside the frameworks listed in Table \ref{tbl:platforms} further solutions exist. {\tt Pyston}\footnote{\url{https://github.com/dropbox/pyston}} is a very new project under current development at Dropbox and is also built on the {\tt LLVM} compiler infrastructure. Further frameworks are topic to research projects such as {\tt Pythran}\cite{guelton2013, guelton2014}, a static {\tt Python} compiler and {\tt parakeet}\cite{rubinsteyn2012}, which follows a similar approach as \hope.

\begin{table}[t]
\caption{Decription of existing packages with the version number used for the benchmarks.}
\begin{center}
\begin{tabular}{l|p{10cm}|c}
Name & Description & Version \\
\hline
\hline
{\tt Numba} & {\tt Numba} is an open source package, which brings JIT compiling to {\tt Python} by generating machine code using the {\tt LLVM} software stack. {\tt Numba} has been targeted to integrate into scientific {\tt Python} applications. Similarly to \hope the user can use a simple decorator to instruct {\tt Numba} to compile a function. As JIT compiler {\tt Numba} is able to use the type information available at runtime to generate the byte code. & 0.13.3 \\ \hline

{\tt Cython} & {\tt Cython} is a compiler for {\tt Python} and for the {\tt Cython} programming language. It allows the user to integrate {\tt C} or {\tt C++} function into {\tt Python} code as well as to compile {\tt Python} code into machine code by adding static type declaration. This is done by using function decorators and type definition extensions. {\tt Cython} supports the functionality of the {\tt NumPy} package. & 0.20.2 \\ \hline

{\tt PyPy} & {\tt PyPy} is a {\tt Python} interpreter written in a restricted subset of the {\tt Python} language itself ({\tt RPython}). {\tt PyPy} is targeted towards speed and efficiency. The included JIT compiler allows translating the {\tt Python} code into native code at runtime. & 2.3.1 \\ \hline

{\tt Nuitka} & {\tt Nuitka} is a static {\tt Python} compiler aiming to support the entire {\tt Python} language definition. It allows the compilation of a {\tt Python} code into an executable. & 0.5.3  \\ \hline

{\tt numexpr} & The {\tt numexpr} package is designed for the efficient evaluation of numerical operations on array-like objects of the {\tt NumPy} package. Additionally, the evaluation can be executed in parallel. &  2.4 \\ \hline

\end{tabular}
\end{center}
\label{tbl:platforms}
\end{table}

\section{HOPE - A Python just-in-time compiler}
\label{sec:architecture}

\hope is a specialized method-at-a-time JIT compiler written in {\tt Python}. It translates {\tt Python} source code into {\tt C++} and compiles the generated code at runtime. In contrast to other existing JIT compliers, which are designed for general purpose, we have focused our development of the subset of the {\tt Python} language that is most relevant for astrophysical calculations. By concentrating on this subset, \hope is able to achieve very high performance for these applications.

\hope is able to translate commonly used unary, binary and comparison operators as well as augmented assign statements. Currently the supported native built-in data types are {\tt bool}, {\tt int} and {\tt float} and their corresponding {\tt NumPy} data types (e.g. {\tt int32}, {\tt float32} resp. {\tt int64}, {\tt float64} etc). This applies to scalar values as well as the {\tt NumPy} arrays with these types. Several features of {\tt NumPy} and mathematical functions like $sin$, $cos$, $exp$ etc. are also supported. HOPE, for instance, allows operations on {\tt NumPy} arrays with the common slicing syntax (e.g. a[5:, 5:] = x). A full list of the functionality backed by the package is published online\footnote{\url{http://pythonhosted.org/hope/lang.html}} and a set of examples is provided on GitHub\footnote{\url{https://github.com/cosmo-ethz/hope/tree/master/examples}}. The package can be used with {\tt Python} 2.7 as well as with {\tt Python} 3.3+. We have tested \hope in Linux and Mac OSX environment with {\tt gcc} and {\tt clang} compilers.

\hope infers the data types by analysing the function signature and inspecting the Abstract Syntax Tree (AST) of the function at runtime. This information is used to translate the {\tt Python} function into statically typed {\tt C++} code. Using {\tt Python}'s native extensions for C API\footnote{\url{https://docs.python.org/2/extending/extending.html}} the code is compiled into a shared object library, which is then loaded into the interpreter runtime. During the process of generating the {\tt C++} code, \hope is able to apply various numerical optimisations in order to improve the execution performance of the compiled function (see section \ref{sec:optimization}). Once compiled, the function is cached to disk to minimise lag on future executions of the same code.

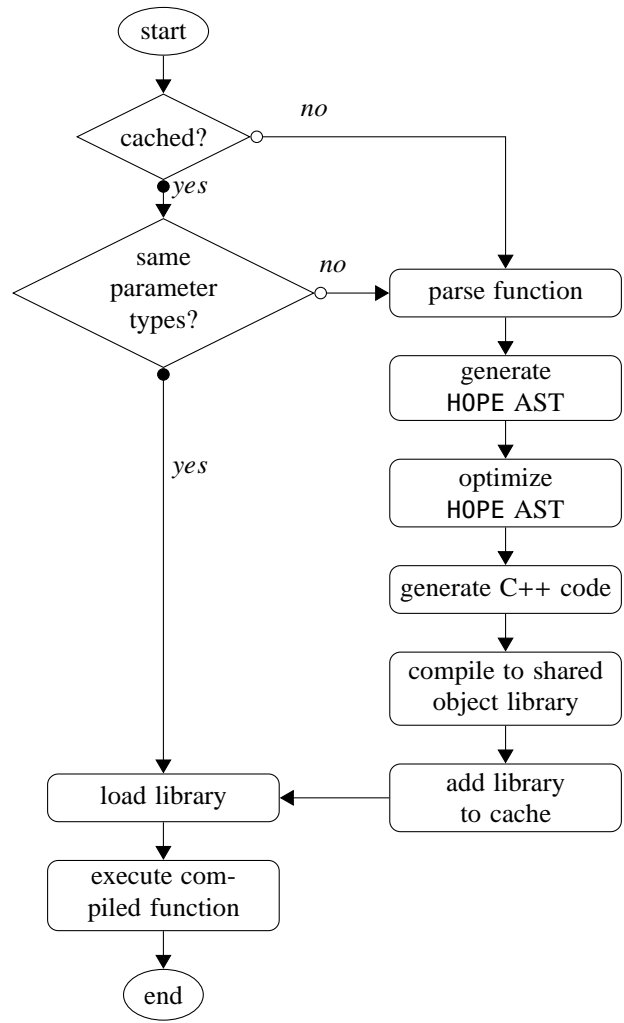
\begin{figure}[h]
\center

\begin{tikzpicture}[%
    >=triangle 60,              
    start chain=going below,    
    node distance=5mm and 45mm, 
    every join/.style={norm},   
    ]
\tikzset{
  base/.style={draw, on chain, on grid, align=center, minimum height=4ex},
  proc/.style={base, rectangle, text width=8em},
  decision/.style={base, diamond, aspect=2, text width=4.5em, inner sep=1pt},
  cloud/.style={base, ellipse},
  block/.style={proc, rounded corners},
  coord/.style={coordinate, on chain, on grid, node distance=5mm and 45mm},
  nmark/.style={draw, cyan, circle, font={\sffamily\bfseries}},
  norm/.style={->, draw},
}

    \node [cloud] (start) {start};
    \node [decision, join] (cached) {cached?};
    \node [decision] (paramtypes) {same parameter types?};

    \node [block, right=of paramtypes] (pythonAst) {parse function};
    \node [block, join] (hopeAST) {generate \hope AST};
    \node [block, join] (optimization) {optimize \hope AST};
    \node [block, join] (generateCPP) {generate C++ code};
    \node [block, join] (compile) {compile to shared object library};
    \node [block, join] (cache) {add library to cache};

    \node [block, left=of cache] (load) {load library};
    \node [block, join] (exec) {execute compiled function};
    \node [cloud, join] (end) {end};

	\node [coord, right= of cached] (c1)  {}; \cmark{1}
	
	\path (cached.south) to node [near start, xshift=1em] {$yes$} (paramtypes);
  		\draw [*->] (cached.south) -- (paramtypes);
	\path (paramtypes.south) to node [near start, xshift=1em] {$yes$} (load);
  		\draw [*->] (paramtypes.south) -- (load);
	   
	\path (paramtypes.east) to node [near start, yshift=1em] {$no$} (pythonAst); 
	  	\draw [o->] (paramtypes.east) -- (pythonAst);

	\path (cache.west) to node [near start, yshift=1em] {} (load); 
	  	\draw [->] (cache.west) -- (load);
	  	
 	\path (cached.east) to node [near start, yshift=1em] {$no$} (c1); 
 		\draw [o->] (cached.east) -- (c1) -- (pythonAst);

\end{tikzpicture}

\caption{Flow diagram of the step executed during a function call decorated with \hope}
\label{fig:flowdiagram}
\end{figure}

We have chosen to generate {\tt C++} code and compiling the code into a library over other approaches, such as the direct generation of byte code. This design decision was made as the intermediate product - the generated code - is human readable, which greatly simplifies the development and debugging process and, furthermore, allows the use of automatic hardware specific optimisation of modern compilers without additional effort. The just-in-time compiling process is described in Fig. \ref{fig:flowdiagram}. A function call undergoes the following several steps:

\begin{description}

\item[Start] The {\tt Python} interpreter loads a function or method previously decorated with the {\tt @hope.jit} decorator.

\item[Cache verification] \hope checks if a compiled version of the requested functions has previously been cached. In case the code is executed the first time, \hope returns a wrapper function containing a reference to the original function.

\item[Parse function] The first time the decorated function is called, the wrapper generates an abstract syntax tree (AST) by parsing the function definition using the {\tt Python} built-in {\tt ast} package.

\item[Generate HOPE AST] Using the visitor pattern, the {\tt Python} AST is traversed and a corresponding \hope specific AST is generated. During the traversal we use the {\tt Python} built-in {\tt inspect} package to infer the data types of the live objects such as parameters, variable and return values. Using this information we are able to statically type the \hope AST i.e. scalar variables will be assigned a type and array liked variable will receive a data type and a shape information. Operations on arrays with the same shape will be grouped into blocks. Furthermore a scope is assigned to each variable in order to identify if it will be passed as parameter or if it has to be instantiated. In the latter case \hope distinguishes between temporary variables, used only once (block scope), and variables used multiple time in the function (body scope).

\item[Numerical optimization] \hope traverses the new AST in order to identify numerical optimisation possibilities and alters the tree accordingly. A detailed explanation of the optimization is discussed in section \ref{sec:optimization}.

\item[Generate C++ code] A {\tt C++} code is generated from the \hope AST. First the function signature is created using the name of the function, the type of the return value and the names of the parameters including their type and shape information. If necessary, statements for the instantiation of new variables are generated. Next, each block is turned into a loop statement according to the shape information of the contained arrays. By grouping array operations, we are able to evaluate the operation element-wise, which improves cache locality. For variables with block scope we can avoid allocating a whole array and instead a scalar value can be allocated (an example is discussed in section \ref{sec:example}). Finally, the generated code is augmented with Python's native extensions API statements, so that the code can be called from the {\tt Python} interpreter.

\item[Compile code to shared object library] The {\tt Python} built-in {\tt setuptools} package is then used to compile a shared object library from the generated code. Using this package and defining the generated code as an extension greatly simplifies the handling of compiler and ensures compatibility with the interpreter. 

\item[Add library to cache] Using the extracted information from the function signature and a hash over the function body the compiled shared object library is cached for future calls. 

\item[Load library] The shared object library is dynamically loaded into the runtime and the pointer to the wrapper is replaced with the pointer to the function in the shared object library to avoid unnecessary overhead.

\item[Execute compiled function] A call to the function is directed to the function in the shared object library and executed with the passed parameters.

\item[Subsequent function call] \hope analyses the types of the passed arguments and queries the cache for a function matching the requested name and arguments. If the system registers a cache hit for a function, the shared object library is then loaded into the runtime and the compiled function is evaluated, otherwise a new function will be generated and compiled.

\end{description}

\subsection{Optimisation}
\label{sec:optimization}

After the \hope specific AST has been created the package performs a static recursive analysis of the expressions to introduce numerical optimisation. The supported possibilities are divided into three groups: 1) simplification of expressions, 2) factorising out subexpressions and 3) replacing the {\tt pow} function for integer exponents. To simplify expression we have used the {\tt SymPy} library\cite{sympy}. {\tt SymPy} is a {\tt Python} library for symbolic mathematics and has been entirely written in Python. To apply the optimization, the AST expression is translated into {\tt SymPy} syntax AST and passed to the {\tt simplify} function. The function applies various different heuristics to reduce the complexity of the passed expression. The simplification is not exactly defined and varies depending on the input. For instance, one example of simplification is that $sin(x)^2 + cos(x)^2$ will be simplified to $1$. 
Furthermore the {\tt SymPy} library is used to factorise out recurring subexpression (common subexpression elimination) using the previously created {\tt SymPy} AST and {\tt SymPy}'s {\tt cse} function. 

From {\tt C++11} on, the {\tt pow} function in the C standard library is not overloaded for integer exponents\footnote{\url{http://en.cppreference.com/w/cpp/numeric/math/pow}}. The internal implementation of the computation of a base to the power of a double exponent is typically done using a series expansion, though this may vary depending on the compiler and hardware architecture. Generally this is efficient for double exponents but not necessarily for integer exponents. \hope therefore tries to identify power expressions with integer exponents and factorizes the expression into several multiplications e.g. $y=x^5$ will be decomposed into $x_2=x^2$ and $y=x_2\times x_2 \times x$. This reduces the computational costs and increases the performance of the execution.

\subsection{Example}
\label{sec:example}

In this section, we show the translation process of \hope using the following simple example:

\begin{lstlisting}[frame=single]
@hope.jit
def fkt(x,y):
    z = x**2 + 1
    return z + y
\end{lstlisting}

As soon as this function is called, \hope will create a statically typed AST using the type information available at runtime. Assuming that the function call was done with {\tt x} and {\tt y} defined as one dimensional {\tt float64} {\tt NumPy} array with same length, the resulting AST can be visualised as the following pseudo code:

\begin{lstlisting}[frame=single]
fkt(numpy::float64[] x, numpy::float64[] y) -> numpy::float64[] {
	new numpy::float64[] ret = numpy::float64[len(x)]
	for(int i = 0; i < len(x); ++i){
		numpy::float64 tmp = x[i] * x[i]
		numpy::float64 z = tmp + 1
		ret[i] = z + y[i]
	}
	return ret
\end{lstlisting}

Using the inspected data types of the parameters and by analysing the mathematical expressions, \hope is able to identify that operations are performed on arrays with compatible data types and dimensions and will group those expressions into a block. The result of this block is a new array with the same data type and dimension, which will also be used as return value. This information tells \hope that a new array has to be instantiated and how the function signature will be defined. Next, the block is turned into a loop over every element of the arrays, which includes the power operation as well as the addition operations. Finally, the power operation on the {\tt x} value with the integer exponent is optimised into a multiplication of the {\tt x} value.

\subsection{Quality assurance}

The \hope package has been developed using the test-driven development (TDD) approach, allowing us to ensure a high level of code quality and numerical accuracy up to type-precision. For every supported language feature we have written a unit test that is executed for all the supported data types and array shape combinations. This results in over 1600 unit tests being executed on our continuous integration server per supported {\tt Python} version.

\section{Benchmarks}
\label{sec:benchmarks}

In this section, we describe benchmarks aimed at assessing the performance of \hope and at comparing it to the other packages described in section~\ref{sec:frameworks}. Since a standardised set of benchmarks for testing the performance of software does not exist, we have generated our own series of tests. The first set are three numerical micro-benchmarks published on the {\tt Julia} language website\footnote{\url{http://julialang.org}} and which are supported by the current version of \hope. These benchmarks have already been applied to various languages. We omitted benchmarks with matrix manipulations, as good performance typically depend on the proper use of specialised matrix libraries such as Intel's Math Kernel Library (MKL) or the Basic Linear Algebra Subprograms (BLAS). Additionally, we defined two benchmarks favoring {\tt numexpr} as they test the ability to compute polynomials and mathematical expression which can be simplified. Finally, we have specified two special purpose benchmarks representing simplified versions of common problems in astrophysics to compare the performance of the general-purpose frameworks. To have a reference and baseline, we additionally implemented all the benchmarks in {\tt C++}. 

For our benchmark tests we have chosen packages where the user should, in principle, be able to write a code in Python and that only minor modifications would needed to enable the performance imporvements. All the benchmarks have been made available online as {\tt IPython} notebooks\footnote{\url{http://hope.phys.ethz.ch}}. The following describes the benchmarks and provide the associated {\tt Python} source code.

\subsection{Fibonacci sequence}

The Fibonacci sequence is a common measurement of the execution speed of repeated recursive function calls. The sequence is defined as:

\begin{equation*}
	F_{n} = F_{n-1} + F_{n-2}, F_{0}=0, F_{1}=1.
\end{equation*}

We executed all the benchmarks with $n=20$ resulting in the answer $6765$. The {\tt Python} implementation is as follows:

\begin{lstlisting}[frame=single]
def fib(n):
    if n < 2:
        return n
    return fib(n - 1) + fib(n - 2)

\end{lstlisting}

\subsection{Quicksort}

Quicksort is a powerful yet simple sorting algorithm that gained widespread adoption in {\tt Unix} as the default sorting function and in the {\tt C} standard library. It extends the complexity of the previous recursive benchmark with a loop, multiple comparison and different data types. The algorithm was implemented as follows:  

\begin{lstlisting}[frame=single]
def qsort_kernel(a, lo, hi):
    i = lo
    j = hi
    if False: return a
    while i < hi:
        pivot = a[(lo+hi) // 2]
        while i <= j:
            while a[i] < pivot:
                i += 1
            while a[j] > pivot:
                j -= 1
            if i <= j:
                tmp = a[i]
                a[i] = a[j]
                a[j] = tmp
                i += 1
                j -= 1
        if lo < j:
            qsort_kernel(a, lo, j)
        lo = i
        j = hi
    return a
\end{lstlisting}

The algorithm is used in the benchmarks to sort an array of 5'000 uniform random {\tt float64}.

\subsection{Pi sum}

This benchmark is a simple approximation of $\pi/2$, testing the run time behavior of two nested for-loops.

\begin{equation*}
	p=\sum\limits_{j=1}^{501}\sum\limits_{k=1}^{10001} \frac{1}{k*k}
\end{equation*}

\begin{lstlisting}[frame=single]
def pisum():
    for j in range(1, 501):
        sum = 0.0
        for k in range(1, 10001):
            sum += 1.0/(k*k)
    return sum
\end{lstlisting}

\subsection{10$^{th}$ order polynomial}

In this numerical benchmark a logarithm is approximated using a Taylor expansion resulting in a 10$^{th}$ order polynomial. The approximation is defined as follows:

\begin{equation*}
	ln(x)\approx\sum\limits_{i=1}^{9} = (-1)^{i-1}\frac{(x-1)^{i}}{i}
\end{equation*}

and can be implemented in {\tt Python} as such:

\begin{lstlisting}[frame=single]
def ln_python(X):
	return (X-1) - (X-1)**2 / 2 + (X-1)**3 / 3 - (X-1)**4 / 4 + 
				   (X-1)**5 / 5 - (X-1)**6 / 6 + (X-1)**7 / 7 - 
				   (X-1)**8 / 8 + (X-1)**9 / 9
\end{lstlisting}

For our benchmarks, a slightly optimized version of this implementation has been used:

\begin{lstlisting}[frame=single]
def ln_python_exp(X):
    x = (X - 1)
    x2 = x*x
    x4 = x2*x2
    x6 = x4*x2
    x8 = x4*x4
    return x - x2 / 2 + x * x2 / 3 - x4 / 4 + x * x4 / 5 - x6 / 6 + x6 * x / 7 - x8 / 8 + x8 * x / 9
\end{lstlisting}

where {\tt X} as been defined as an array of 10'000 uniform random {\tt float64}. 

\subsection{Simplify}

This benchmark has been specified as

\begin{equation*}
	y(x)=sin^2(x) + \frac{(x^3+x^2-x-1)}{(x^2 + 2 * x + 1)} + cos^2(x)
\end{equation*}

\begin{lstlisting}[frame=single]
def y(X):
	return np.sin(x)**2 + (x**3 + x**2 - x - 1)/(x**2 + 2*x + 1) + np.cos(x)**2
\end{lstlisting}

Where {\tt X} as been defined as  an array of 5'000 uniform random {\tt float64}. The benchmark tests the ability of the packages to efficiently compute polynomial expressions. As the expression can be simplified to $y=x$ this benchmark will favour frameworks, which analyse and optimise the expression such as {\tt numexpr} and \hope. 

\subsection{Pairwise distance}

Computing the two-point correlation function for a set of points is a common problem in many areas of astronomy and cosmology. A similar but simplified problem is the computation of the distance between all the points. Using the $l^2$ norm this results in a $n \times n $ matrix for a given input array of size $n$. The distance is defined as:

\begin{equation*}
	D_{i,j} = \sqrt{\sum\limits_{k=0}^{N} (X_{i,k} - X_{j,k})^{2}}, \forall i \in I, \forall j \in J
\end{equation*}

where $N=3$ is the number of dimensions and $I=J=1000$ is the number of points in the array. The naive solution used in the benchmark is implemented as follows:

\begin{lstlisting}[frame=single]
def pairwise_python(X, D):
    M = X.shape[0]
    N = X.shape[1]
    for i in range(M):
        for j in range(M):
            d = 0.0
            for k in range(N):
                tmp = X[i, k] - X[j, k]
                d += tmp * tmp
            D[i, j] = np.sqrt(d)
\end{lstlisting}

Additionally we benchmark the vectorized implementation of this test using the {\tt NumPy} package:

\begin{lstlisting}[frame=single]
def pairwise_numpy(X, D):
    M = X.shape[0]
    for i in range(M):
        D[i, :] = np.sqrt(np.sum((X[i, :] - X[:]) ** 2, axis=1))
\end{lstlisting}

\subsection{Point Spread Function}

This benchmark has been inspired by calculations that simulate astronomical imaging data. One step in these simulations is to make realistic simulations of the images of stars. For many astronomical applications stars are smaller than the resolution of the instrument. Therefore, the resulting images are realisations of the Point Spread Function (PSF) comming from the finite resolution and atmospheric effects. A good model for the PSF of ground-based telescopes is a circular Moffat profile \cite{Moffat1969}, given by

\begin{equation*}
	I(r)=I_{0}\Bigg[1+\Big(\frac{r}{\alpha}\Big)^{2} \Bigg]^{-\beta}
\end{equation*}

where $I_{0}$ is the value at the origin (r=0), $\alpha$ and $\beta$ are parameters depending on the conditions of the observation (see Appendix B in \cite{berge2013}). The numerical integration in the x and y direction is done using Gauss-Legendre integration with 7$^{th}$ order coefficients:

\begin{lstlisting}[frame=single]
def pdf(density, x_range, y_range, x_center, y_center, w2D, r50, b, a):
    for x in range(x_range):
        for y in range(y_range):
            dr = np.sqrt((x - x_center) ** 2 + (y - y_center) ** 2)
            density[x, y] = np.sum(w2D * 2 * (b - 1) / (2 * np.pi * (r50 * a)**2) * (1 + (dr / (r50 * a))**2)**(-b))
            return density
\end{lstlisting}

The {\tt density} and {\tt w2D} variables are $20 \times 20$ and $7\times7$ {\tt Numpy} arrays, respectively, while the other parameter are defined as scalar values. A very similar approach has been implemented in the {\tt UFig} project \cite{berge2013}.

\section{Results}
\label{sec:results}

We ran the benchmarks on a MacBook Pro OS X 10.9.2 with a Intel Core i7 2.3 Ghz processor and 16 GB 1600 MHz DDR3 memory as well as on a Mac Pro (see \ref{sec:macproresults} for details). The results with the second system can be found in the \ref{sec:macproresults}. Table \ref{tbl:platforms} shows the versions of the packages used to conduct the test runs. To compile the {\tt C++} code and \hope benchmarks we used {\tt clang-503.0.40} and the {\tt -march=native}, {\tt -stdlib=libc++} and {\tt -std=c++11} compiler options. 

The benchmarks from section \ref{sec:benchmarks} have been executed with all the frameworks of Table \ref{tbl:platforms} except {\tt numexpr} which could only be used for the computation of the 10$^{th}$order polynomial and the simplify benchmark. Every test has been executed 100 times using the built-in {\tt timeit} package and the median of the measured runtime has been used for comparison in order to reduce the influence of other processes interfering with the timing accuracy.

For all the benchmarks using {\tt Cython} we have disabled the {\tt wraparound} and {\tt boundscheck} compiler flags. The benchmarks with {\tt numexpr} have been executed with parallelization using the 8 cores available on the testing infrastructure. Table \ref{tbl:benchmarksresults} shows the benchmark run times relative to the {\tt C++} implementation. The Pairwise distance benchmark has been performed using both, the plain {\tt Python} implementation, as well as the vectorized implementation using the {\tt NumPy} package (see results in parenthesis in the table).

Generally {\tt Cython} was able to improve the performance by a factor 1.2$\times$ up to 53$\times$ compared to the {\tt Python} runs. An exception is the simplify benchmark where the timing was marginally worse. The runs with the {\tt Numba} package show that, if the framework is able to properly infer the data types and compile the function to {\tt LLVM}, the performance is comparable to {\tt Cython}. We have not been able to perform the Quicksort benchmark using {\tt Numba} as the execution resulted in an {\tt Internal error}. This might be due to the currently limited support for recursive function calls. The Pairwise distance benchmark shows that vectorizing the code with the {\tt NumPy} package can drastically improves performance. The impact is more pronounced on {\tt Python} and {\tt Nuitka} (155$\times$ and 131$\times$, respectively) as with {\tt PyPy} and {\tt NumPy} (4.6$\times$). However, vectorisation is not always possible or can be difficult and may increase memory consumption due to vector broadcasting. The alternative interpreter {\tt PyPy} was able to improve the performance compared to {\tt CPython} in the first three benchmarks (1.7$\times$ - 26$\times$) but performed worse in the others (1.2$\times$ - 17$\times$ slower). {\tt Numexpr} achieved better runtimes than {\tt Python} for the two benchmarks conducted. The improvement of the parallelisation was small, which may be ascribable to the parallelisation overhead. For larger problems the speed up could be larger. Attempts to conduct the benchmarks with the {\tt parakeet} project, which is implementing a similar approach to \hope, succeeded only for the Simplify and Pairwise distance benchmark. The improvement for the first benchmark was small and for the second, {\tt parakeet} was faster than the vectorized {\tt NumPy} implementation by a factor of 5.1$\times$.

Our \hope package was able to speed up the computation in all of the benchmarks. A very large improvement can be seen in the pairwise distance benchmark. This can be ascribed to the naive {\tt Python} implementation using multiple nested loops. As the bad performance of loops in {\tt Python} is commonly known, we expect developers to implement this benchmark using {\tt Numpy} instead. Therefore we disregard this speed up for the further comparisons. As can be seen in Table \ref{tbl:benchmarksresults} \hope improved the performance by a factor of 2.4$\times$ to 119$\times$ compared to the {\tt Python} runs and was only marginally slower than the native {\tt C++} implementation. 

For the Pi sum benchmark {\tt Numba} and {\tt PyPy} were able to improve the performance as much as \hope and to match that of {\tt C++}. In the Pairwise distance benchmark {\tt Cython} outperformed \hope and reached the same performance as the native {\tt C++} implementation. The simplify benchmark shows the power of the optimisation capabilities of the \hope package. Disabling the optimisation option in \hope would result in similar timing as the native {\tt C++} implementation. For specialised problems such as the Point Spread Function, \hope and {\tt C++} both clearly outperform all the alternative frameworks tested. The benchmarks conducted on the Mac Pro system yield comparable results (see Table \ref{tbl:benchmarksresultspro} of \ref{sec:macproresults}).

\begin{table}[tdp]
\caption{Benchmarks times relative to {\tt C++}. Best results are highlighted in bold.}
\begin{center}
\begin{minipage}{.9\textwidth}
\begin{center}
\begin{tabular}{l|c|c|c|c|c|c|c|c|c}
& \pbox{20cm}{{\tt Python} \\ ({\tt NumPy})} & {\tt Numba} & {\tt Cython} & \pbox{20cm}{{\tt Nuitka} \\ ({\tt NumPy})} & \pbox{20cm}{{\tt PyPy} \\ ({\tt NumPy})} & \pbox{20cm}{{\tt numexpr} \\ (8 cores)} & \hope & {\tt C++}\\
\hline
\hline
Fibonacci & 57.4 & 65.7\footnote{\label{nollvm}	{\tt Numba} was not able to compile down to LLVM} & 1.1 & 26.7 & 21.1 & --- & 1.1 & \bf{1.0} \\ \hline 
Quicksort & 79.4 & ---\footnote{\label{error}	Compilation attempt resulted in {\tt Internal error}} & 4.6 & 61.0 & 45.8 & --- & 1.1 & \bf{1.0} \\ \hline 
Pi sum & 27.2 & \bf{1.0} & 1.1 & 13.0 & \bf{1.0} & --- & \bf{1.0} & \bf{1.0} \\ \hline
10$^{th}$order & 2.6 & 2.2 & 2.1 & 1.2 & 12.1 & 1.4 & 1.1 & \bf{1.0} \\ \hline
Simplify & 1.4 & 1.5\footref{nollvm}\footref{error} & 1.8 & 1.4 & 23.2 & 0.6 & \bf{0.015} & 1.0 \\ \hline
\pbox{20cm}{Pairwise \\ distance} & \pbox{20cm}{1357.8 \\ (8.7)} & 1.8 & \bf{1.0} & \pbox{20cm}{1247.7 \\ (9.5)} & \pbox{20cm}{277.8 \\ (60.4)} & --- & 1.7 & \bf{1.0} \\ \hline
Star PSF & 265.4 & 250.4\footref{nollvm} & 46.2 & 234.6 & 339.5 & --- & 2.2 & \bf{1.0} \\ \hline
\end{tabular}

\end{center}
\end{minipage}
\end{center}
\label{tbl:benchmarksresults}
\end{table}%

\section{Discussion}
\label{sec:discussion}

When using {\tt Cython}, the user has to provide to the package a statically typed function signature, as well as statically typed variables in order to achieve the desired performance improvements. The package {\tt numexpr} yields good performance (as can be seen in  Table \ref{tbl:benchmarksresults}) but it requires the user to write expressions as strings. This limits the applicability and, more importantly, removes syntax highlighting and variable recognition in editors and integrated development environments (IDE). The benchmarks indicate that the {\tt PyPy} interpreter is only able to partially increase the performance compared to {\tt CPython}. The project is under active development and interesting concepts are being addressed within it. Packages with C extension such as {\tt SciPy} are currently not fully supported, which limits the use of {\tt PyPy} in scientific applications. Since compiling the {\tt Python} code with the static compiler {\tt Nuitka} improved the performance only marginally, the overhead arising from the compilation process do not appear to be justified. {\tt Numba}, which is the only package besides \hope that does not require the user to alter the code or change the runtime environment, shows good performance as soon the package is able to compile down to {\tt LLVM}. As the project is also under active development we expect that further support and features will be implemented soon. Focusing on a subset of the {\tt Pyhton} language enables \hope to generate C++ code targeted towards high execution performance without the need for the user to modify the Python implementation. The performance differences compared to the {\tt C++} implementation arise through small overheads introduced by the code generation process. It has to be noted that \hope is still under active development and many language features of {\tt Python} are currently not supported. In cases where \hope is not able to translate the code it will provide the user the according information including the line of code, which caused the problem. In these cases exploring possible improvements in performance through {\tt Cython} would be an option. This requires the user manually adapt the code. For an experienced user gains are likely to be possible.

\section{Conclusion}
\label{sec:conclusion}

{\tt Python} is becoming increasingly popular in the science community due to the large variety of freely available packages and the simplicity and versatility of the language.  However, a drawback of {\tt Python} is the low runtime and execution performance of the language. For many use cases this is acceptable but for large simulations and numerical computations, such as those used in astrophysics and cosmology, accelerating the performance of codes is crucial. This can be achieved by parallelizing the computation on multicore CPU architectures. Alternatively and complementarily, the single thread performance of the code can be optimized. Rewriting the application in C (or other compiled languages) can be time consuming and reduces the readability and maintainability of the code. A set of solutions exists in the {\tt Python} landscape to improve the performance, such as alternative interpreters, static {\tt Python} code compiler or just-in-time compilers. We find that some those solution are intrusive i.e. they require the user to change the code and some are not able to fully achieve the speed of a corresponding {\tt C/C++} implementation.

To address these limitations, we introduced \hope, a specialised {\tt Python} just-in-time compiler able to apply numerical optimisation to mathematical expressions during the compilation process. We conducted different benchmarks to assess its performance and compared it with existing solutions. The tests show that \hope is able to improve the performance compared to plain {\tt Python} by a factor of 2.4$\times$ - 119$\times$ depending on the benchmark scenario. We find that the performance of our package is comparable to that of {\tt C++}. Some of the other packages that we tested are also able to improve the execution speed but do not increase the performance in specialized test cases such as the computation of a ground-based  point spread function. We have used our package to improve the performance of the {\tt PyCosmo} project \cite{refregier2014} as well as to be able to rewrite the Ultra fast image generator ({\tt UFig}) {\tt C++} package\cite{berge2013} in {\tt Python} without compromising its performance. We plan to apply \hope to further projects and therefore continuously increase its supported language features and improve its optimization capabilities. To simplify the installation we are distributing the code through the central PyPI server\footnote{\url{https://pypi.python.org/pypi/hope}} and provide the full documentation of the package online\footnote{\url{http://hope.phys.ethz.ch}}. In \ref{sec:distribution} we describe the distribution and installation of \hope.

%
%

\newpage

\appendix

\section{Distribution}
\label{sec:distribution}

Detailed documentation, supported language subset and installation instructions can be found on the package website \url{http://hope.phys.ethz.ch}. The \hope package is released under the GPLv3 license and has been uploaded to PyPI\footnote{\url{https://pypi.python.org/pypi/hope}} and can be installed using pip\footnote{\url{www.pip-installer.org/}}: 

\begin{verbatim}
$ pip install hope --user
\end{verbatim}

This will install the package and all of the required dependencies. The development is coordinated on GitHub \url{http://github.com/cosmo-ethz/hope} and contributions are welcome.

\section{Mac Pro results}
\label{sec:macproresults}

The benchmarks described in section \ref{sec:benchmarks} have been conducted on a Mac Pro system with an Intel Xeon E5 Ivy Bridge 12-core (2.7 GHz) and 64GB (4x16GB) of 1866MHz DDR3 memory to test the effect of different hardware. As can be seen in Table \ref{tbl:benchmarksresultspro}, the timings are comparable to the results in Table \ref{tbl:benchmarksresults}, leading to the same conclusions as discussed in section \ref{sec:discussion}.

\begin{table}[h]
\caption{Benchmarks times relative to {\tt C++} (Mac Pro). Best results are highlighted in bold.}
\begin{center}
\begin{minipage}{.9\textwidth}
\begin{center}
\begin{tabular}{l|c|c|c|c|c|c|c|c|c}
& \pbox{20cm}{{\tt Python} \\ ({\tt NumPy})} & {\tt Numba} & {\tt Cython} & \pbox{20cm}{{\tt Nuitka} \\ ({\tt NumPy})} & \pbox{20cm}{{\tt PyPy} \\ ({\tt NumPy})} & \pbox{20cm}{{\tt numexpr} \\ (8 cores)} & \hope & {\tt C++}\\
\hline
\hline
Fibonacci & 65.5 & 62.6\footnote{\label{nollvm}	{\tt Numba} was not able to compile down to LLVM} & \bf{1.0} & 31.6 & 12.4 & --- & \bf{1.0} & \bf{1.0} \\ \hline 
Quicksort & 109.0 & ---\footnote{\label{error}	Compilation attempt resulted in {\tt Internal error}} & 5.0 & 62.7 & 39.2 & --- & 1.1 & \bf{1.0} \\ \hline 
Pi sum & 31.9 & \bf{1.0} & \bf{1.0} & 16.6 & 1.1 & --- & \bf{1.0} & \bf{1.0} \\ \hline
10$^{th}$order & 4.0 & 3.9 & 3.8 & 3.5 & 25.8 & 2.3 & 1.2 & \bf{1.0} \\ \hline
Simplify & 1.4 & 1.3\footref{nollvm}\footref{error} & 1.4 & 1.2 & 7.9 & 0.8 & \bf{0.017} & 1.0 \\ \hline
\pbox{20cm}{Pairwise \\ distance} & \pbox{20cm}{2076.6 \\ (13.9)} & 16.2 & \bf{1.0} & \pbox{20cm}{1460.8 \\ (13.7)} & \pbox{20cm}{254.4 \\ (75.2)} & --- & 1.7 & \bf{1.0} \\ \hline
Star PSF & 374.7 & 385.9\footref{nollvm} & 98.8 & 348.7 & 500.1 & --- & 2.0 & \bf{1.0} \\ \hline
\end{tabular}
\end{center}
\end{minipage}
\end{center}
\label{tbl:benchmarksresultspro}
\end{table}%

\newpage

\bibliographystyle{elsarticle-num}

\bibliography{hope}

\end{document}